# Prediction Air Temperature in Geothermal Heat Exchangers Using Pseudorandom Numbers: The New DARL Model

C. Ramírez-Dolores[a,e], J.C. Zamora-Luria[b], J.A. Altamirano-Acosta[c], L. Sarao-Cruz[d], P. Jiménez-Palma[e], J. Moreno-Falconi[f]

[a]Facultad de Ingeniería Industrial. Universidad de Sotavento A.C.
[b]Department of Geoscience. Aarhus University.
[c]Programa de Posgrado en Ingeniería Industrial. Tecnológico Nacional de México Campus La Venta.
[d]Departamento de Ingeniería en Gestión Empresarial. Tecnológico Nacional de México Campus La Venta.
[e]Departamento de Ingeniería Industrial. Tecnológico Nacional de México Campus La Venta.
[f]Escuela Superior de Ingeniería Química e Industrias Extractivas. Instituto Politécnico Nacional.
Coatzacoalcos (Veracruz), México. La Venta (Tabasco), México. Ciudad de México, México. Aarhus, Denmark.
Corresponding author email: cesar.ramirez@us.edu.mx (ORCID: 0000-0002-4817-6254)

*Abstract*—The use of Earth-Air-Water Heat Exchangers (EAWHE) for sustainable air conditioning has not been widely studied. Due to their experimental nature, methods of characterizing internal thermal air distribution impose high dependence on instrumentation by sensors and entail data acquisition and computational costs. This document presents an alternative method that estimates air temperature distribution while minimizing the need for a dense network of sensors in the experimental system. The proposed model, DARL (Data of Air and Random Length), can predict the temperature of air circulating inside EAWHEs. DARL is a significant methodological advance that integrates experimental data from boundary conditions with simulations based on pseudo-random numbers (PRNs). These PRNs are generated using Fermat's prime numbers as seeds to initialize the generator. Ordinary linear regressions and robust statistical validations, including the Shapiro–Wilk test and root mean square error, have demonstrated that the model can estimate the thermal distribution of air at different lengths with a relative error of less than 6.2%. These results demonstrate the model's efficiency, predictive capacity, and potential to reduce dependence on sensors. The model is thus established as a viable alternative for designing, evaluating, and optimizing geothermal systems under real operating conditions.

## I. Introduction (Heading 1)

The growing global demand for energy has increased the pressure to extract and refine fossil resources, which has generated significant environmental impacts, such as greenhouse gas emissions [1, 2]. In this context, transitioning to renewable energy sources and related technologies has become a strategic priority for governments, researchers, and energy system designers. However, in addition to adopting clean aditionally, optimizing energy use in everyday applications is essential, particularly for technologies that are becoming increasingly indispensable for human activities. One example is the thermal conditioning of built spaces, which accounts for a significant proportion of global energy consumption [3].

Geothermal heat exchangers, especially Earth Air Water (EAWHE) types, are among the most promising emerging solutions due to their high thermal performance, low energy consumption, and adaptability to different geological environments [4-7]. These systems use the thermal stability of the subsoil to modify the temperature of the air inside the pipe arrangement via a mechanical system, that uses the ground as a heat source or sink. This enables the air leaving the system to be maintained at comfortable temperatures. Using these systems offers several advantages for air conditioning operations, including energy savings, which impact economic savings; low construction and maintenance costs; and the absence of refrigerants during operation, which is also considered relevant. Implementing EAWHE systems in hot and humid regions has significant advantages due to the high thermal conductivity of sandy soil, the ease with wich the subsoil can be drilled into, and the presence of shallow aquifers in some locations.

Implementing these systems with physical sensors can present technical and economic challenges, especially with experimental systems. Disconnection failures, loss of underground instruments, and prolonged calibration times may affect the reliability of the data, increase operating costs, and endanger the integrity of plastic pipes [8, 9]. In response to these limitations, this study proposes developing a novel numerical model that can predict air temperature distribution in geothermal heat exchangers. One of the most impactful advantages of this model is that it would minimize sensor use and strengthen thermal performance evaluation criteria, in addition to adopting a new strategy that combines experimental parameters with computational methods. The proposed model is based on thermal differences and integrates experimental parameters and synthetic data obtained from simulations using pseudo-random numbers (PRNs), which are generated using Fermat's prime numbers as seeds to initialize the generator [10, 11]. The methodology employs ordinary linear regressions on PRN series to estimate temperatures in different segments of the exchanger (i.e., synthetic data) and validates the results using statistical tests such as the Shapiro-Wilk test and root mean square error (RMSE) test [12, 13].

This experimental and numerical approach allows us to simulate the thermal behavior of air circulating inside the EAWHE under real operating conditions with a relative error of less than 6.2%. This demonstrates its applicability. The model provides an efficient alternative to traditional methods requiring spatial discretization, complex meshes, and substantial computational power. The aim of this research is to develop a new model (DARL), to determine the temperature of air circulating in EAWHEs in shallow aquifers. The model could be a key tool in designing, evaluating, and optimizing EAHWE. It has potential applications in thermal simulators, sustainable air conditioning projects, and heat transfer studies in porous media.

## II. METHODOLOGY

This section describes the experimental design, characterization of the study area, construction of the prototype, and development of the model based on pseudo-random rational numbers (PRNs). It also describes the validation procedure.

### A. Study area and geological conditions

The study was conducted in the city of Coatzacoalcos (18° 08' N latitude; and 94° 27' W longitude) in southern Veracruz (Mexico), a coastal region characterized by a hot-humid climate, sandy soil, and the presence of shallow aquifers [14]. Geologically, the soil in Coatzacoalcos has a high proportion of sand strata, which are used as foundation layers for different structures. The particle type corresponds to coarse grain (mostly) with a tendency toward fine grain depending on depth. The mineralogical characteristics of the soil indicate a content of quartz, feldspars, and, to a lesser extent, lithic fragments [15].

### B. Experimental prototype

A horizontal EAWHE prototype with polyvinyl chloride (PVC) piping with an outer diameter of 76.20 mm was constructed and installed. The straight sections were connected using 90° elbows. The EAWHE's geometry includes a horizontal pipe that is 6 metres long, as well as vertical inlet and outlet pipes. Each of these is 2.50 metres long.

At the EAWHE, six sensors (Pt-100 sensors, IST brand) were strategically installed to record the temperature (measurement uncertainty of ± 0.05 °C) of the air induced by a blower (Figure 1). The following configuration was used to record the temperature: a) Inlet air temperature: the first sensor ($T_{in}$) was placed 10 cm behind the first elbow (air descent section). b) Sensors 1 and 7 are installed 10 cm after and before the elbows, respectively. c) The subsequent sensors (2, 3, and 4) were placed as follows:

Sensor 2 (S2) is located 0.90 meters away from Sensor 1 and 3.30 meters away from Sensor $T_{in}$.

Sensor 3 (S3) is located 1.90 meters away from Sensor 1 and 4.30 meters away from Sensor $T_{in}$.

Sensor 4 (S4) is located 2.90 meters away from Sensor 1 and 5.30 meters away from Sensor $T_{in}$

During the experimental phase, a temperature sensor was ($T_w$) placed inside the groundwater. The sensor was installed submerged and did not make contact with the substrate.

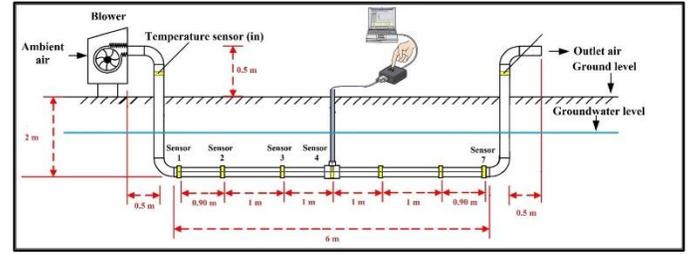

Fig. 1. Schematic diagram of the experimental EAWHE.

### C. Experimental runs

Several experimental runs were carried out under continuous operating conditions to record the EAWHE's thermal behavior. Two representative experiments are reported below to compare the proposed model with the experimental data.

**Experiment A:** a total length of 5.40 meters is considered and the simulated temperature corresponds to that expected at positions S1 (2.50 metres), S2 (3.40 metres) and S3 (4.40 metres). Stable air temperature values are assumed at the inlet of the exchanger and at sensor position 4. Thus, two constants are used: $T_{in}$ = 31.01 °C and S4 = 25.81 °C. The groundwater temperature is also considered: $T_w$= 24.28 ± 0.09 °C. This case was proposed to simulate the air temperature at sensor positions 1, 2, and 3.

Table 1 shows the system configuration to be simulated as simple block diagrams. Yellow rectangles indicate where the proposed model should predict the air temperature. The lengths at which the model was tested are also shown.

TABLE I. SETTINGS FOR PREDICTING TEMPERATURE (EXPERIMENT A).

| Test setup | Length for predicting temperature by configuration (m) |
|---|---|
| $T_{in}$ → S1 (2.50 m) → S4 | 2.50 |
| $T_{in}$ → S1 (3.40 m) → S4 | 3.40 |
| $T_{in}$ → S1 (4.40 m) → S4 | 4.40 |

**Experiment B:** The total length of this experiment is 8.30 m. The simulated temperature corresponds to that expected at lengths of 2.50 m (S1 position), 3.40 m (S2 position), 4.40 m (S3 position), and 5.40 m (S4 position). Stable values were considered for the air temperature at the inlet of the exchanger and at the sensor position ($T_{in}$= 31.01 °C and S7= 24.54 °C), as well as the water temperature ($T_w$= 24.28 ± 0.09 °C).

The experiments were designed to collect real operating data and simulate air temperatures at sensor positions 1, 2, 3, and 4. Table 2 shows the block diagram of the simulated system configuration. The yellow rectangles indicate where the proposed model must predict the air temperature. The length values where the proposed model was tested are also shown.

TABLE II. SETTINGS FOR PREDICTING TEMPERATURE (EXPERIMENT B).

| Test setup | Length for predicting temperature by configuration (m) |
|---|---|
| 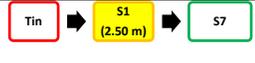 | 2.50 |
| 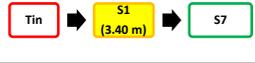 | 3.40 |
| 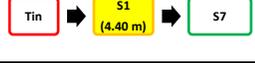 | 4.40 |
| 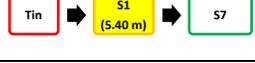 | 5.40 |

In each experiment, temperatures were recorded every seven seconds for 300 minutes. The data were stored in the data acquisition system and analyzed later. Values of Tin, S4, and S7 were considered constant for each configuration, enabling thermal limits to be established for the simulation.

*D. Pseudo-random number generation (PRN)*

We used PRNs to simulate the thermal distribution of air as a function of length in the heat exchanger. The PRNs were generated using the runif function in the RStudio software package. This function is based on precalculated numbers derived from the Mersenne-Twister algorithm, which generates sequences of vectors that are considered to be pseudo-random numbers [16]. PRNs are generated from an initial value and a transformation; according to the principle by which the generator produces the sequence of values (see reference [17], for details of the algorithm). To generate the PRNs using the runif function, the seed values had to be set. Five seeds were chosen, corresponding to Fermat´s prime numbers: 3, 5, 17, 257, and 65537 [11]. The application of Fermat numbers in this type of study has not been documented previously, making this research unprecedented. Each seed generates a sequence of values that are distributed within the interval $T_{min}$ to $T_{max}$.

Where:

$T_{max}$: Air temperature at inlet (Tin).

$T_{min}$: Air temperature at the end (S4 or S7).

n: number of values (equivalent to length).

Example of code used:

```
# SCRIPT: Synthetic Data Generation
# PURPOSE: Generate a sample of N=538 values within a range to simulate temperature
# SECTION 1: REPRODUCIBILITY CONFIGURATION AND DATA GENERATION
# 1.1. Set the Seed
set.seed(3)
# 1.2. Generation of the Synthetic Variable: 538 observations are generated with a uniform distribution (runif)
# in the closed interval [25.81, 31.01].
X1_3 <- runif(538, min = 25.81, max = 31.01)
# SECTION 2: STRUCTURING AND EXPORTING THE DATA FRAME
# 2.1. Creating and sorting the data frame. The data is organized in a single-column data frame and sorted in ascending order.
df_X1_3 <- data.frame(Ordered_Value = sort(X1_3))
# 2.2. Export to CSV format. The sorted data is exported to a CSV file.
# 'row.names = FALSE' prevents the inclusion of the R index column.
write.csv(df_X1_3, "X1_3.csv", row.names = FALSE)
#SECTION 3: REPORTING AND VERIFYING RESULTS
# 3.1. Printing the Data Frame (First Rows). Display the resulting data frame in the console.
cat("\n--- Data Frame Generated  (df_X1_3) ---\n")
print(df_X1_3)
## 3.2. Statistical Summary for Verification
cat("\n--- Statistical Summary (summary) ---\n")
summary(df_X1_3)
## 3.3. Structure Verification. Display the data structure (variable type and number of observations).
cat("\n---  Data Frame Structure (str) ---\n")
str(df_X1_3)
```

*E. Linear regression on PRN*

Each series of generated PRNs was subjected to ordinary linear regression, with the X-axis representing pipe length and the Y-axis representing the PRN values obtained by the generator. Regression equation (1) provides two key parameters.

$$T_\varphi = \alpha + \beta x \quad (1)$$

$T_\varphi$: Temperature obtained by ordinary linear regression from PRN, as a function of length (°C). (Synthetic data).

These values were integrated into the proposed model to calculate the simulated temperature (T) in each segment of the EAWHE. This procedure was performed for each of the seeds used Fermat numbers used as seeds, which was essential in order to define a correlation criterion between the variables and determine the effectiveness of the most effective seed for predicting temperatures.

*F. Validation*

The following statistical tests were applied to evaluate the distribution of the PRNs and assess the models performance:

**Shapiro-Wilk test (α= 0.05):** Used to analyze which generated PRNs complied with the normality assumption, i.e., were distributed in a Gaussian manner [12].

**Root Mean Square Error (RMSE):** The performance of the proposed model was evaluated using RMSE (Equation 2). RMSE corresponds to the standard deviation of prediction errors, which are used to measure the distance of data points from the regression line. RMSE is also considered a measure of the dispersion of these values. In other words, RMSE indicates how concentrated the data are on the best-fit line [13].

$$RMSE = \sqrt{\frac{1}{x}\sum_{i=1}^{x}(Y_i^{obs} - Y_i^{sim})^2} \quad (2)$$

Where:

$RMSE$: Root Mean Square Error.

$x$: Number of measurements.

$Y_i^{obs}$: Observed values (experimental).

$Y_i^{sim}$: Simulated values.

## III. RESULTS

A numerical model was constructed to predict the temperature of the air circulating inside the EAWHE system. This was based on experimental measurements of air temperature at the inlet and outlet, as well as at various intervals, and on groundwater temperature records. The considerations and assumptions underlying this model are presented below.

Considerations:

a) Heat transfer by convection occurs between the air flowing inside the pipe and its inner surface; b) heat transfer by conduction due to the pipes thickness, and c) heat transfer by convection between the groundwater and the pipes outer surface.

Assumptions:

a. Heat transfer is in a steady state.

b. The thermophysical properties of the groundwater and air do not change over time.

c. The velocity of air entering the EAWHE is constant.

d. The pipe has a uniform, circular cross-sectional area.

e. The pipes relative roughness is negligible

The proposed model incorporates temperature parameters obtained through experimentation and two parameters derived computationally. PRNs were used to perform ordinary linear regressions in each simulated case, and the point value of each linear regression ($T_\varphi$) was used independently to be integrated into equation 3, proposed as a model to simulate the temperature in different longitudinal sections of the EAWHE.

$$T = \left[\left(\frac{\frac{T_{max} - T_{min}}{T_\varphi - T_w}}{R^2}\right)(T_w)\right] + T_\varphi \quad (3)$$

Where:

$T$: Simulated air temperature value in the EWAHE in different longitudinal sections (°C).

$T_{max}$: Maximum temperature recorded in this case is the temperature of the air entering the EAWHE (°C).

$T_{min}$: Minimum temperature: the values recorded by Sensor 4 and Sensor 7 are used (°C).

$T_w$: Groundwater temperature measured *in situ* (°C).

$R^2$: Coefficient of determination obtained from ordinary linear regression from PRN.

The proposed model is a correlation function designed to predict the air temperature profile along the EAWHE. This model has been developed under the principle of energy conservation under steady-state assumptions, which implies that the thermophysical properties of the air circulating inside and the groundwater affecting the outside of the EAWHE are considered constant over time. The distinctive feature of the DARL model is its method for generating temperature estimates, which gives it an advantage over common techniques.

### A. Experiment A

Figure 2 shows a graph of the experimental air temperature values recorded by the sensors during the experiment, alongside the values obtained from Equation 3.

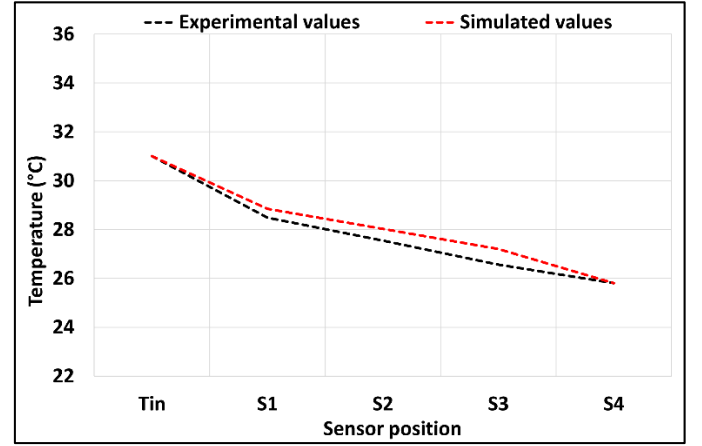

Fig. 2. Experimental and simulated air temperature values in Experiment A.

The dotted red line shows simulated values obtained using the DARL model, while the dotted black line shows the temperature values recorded by the sensors. The figure shows a similar distribution of magnitude as a function of sensor position (length of the EAWHE). Between sensors 2 and 3, the difference between the experimental and simulated values is minimal. The relative errors between the simulated and experimental values were determined. Figure 3 shows the relationship between relative error and EAWHE length for Experiment A.

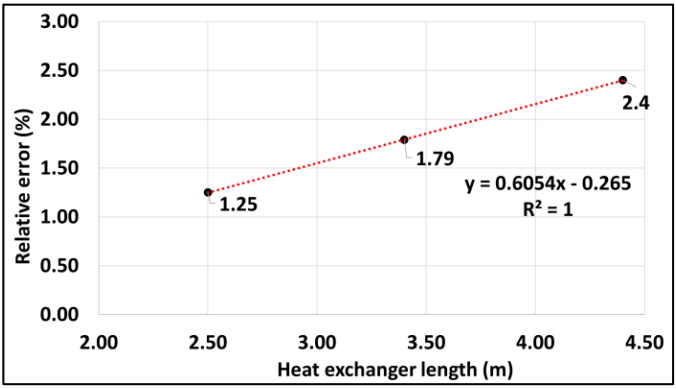

Fig. 3. Estimated relative errors for the conditions of Experiment A.

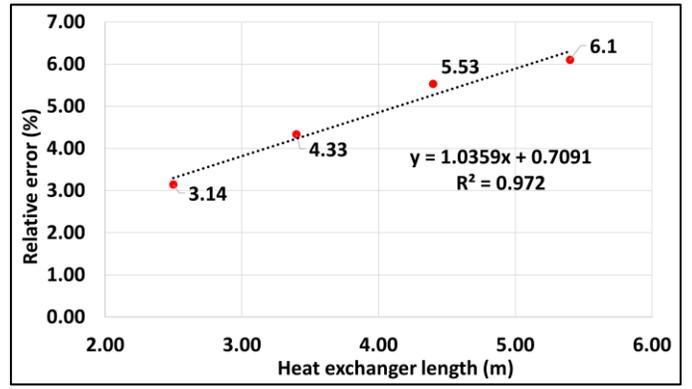

Fig. 5. Estimated relative errors for the conditions of Experiment B.

The proposed model can generate air temperature values with an error of less than 2.50%. For a length of 2.50 meters, the relative error between the experimental and simulated values is 1.25%. For a length of 3.40 meters, the calculated error is 1.79%. For a length of 4.40 meters, the relative error is 2.40%.

## B. Experiment B

Figure 4 shows the air temperature values recorded by the sensors and the results obtained from Equation 3 for the conditions of Experiment B.

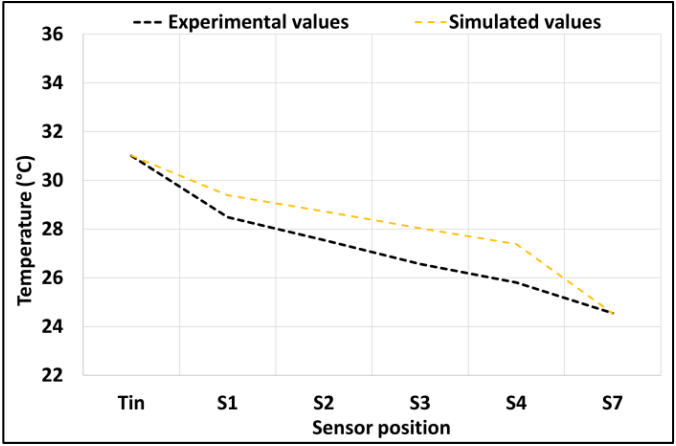

Fig. 4. Estimated relative errors for the conditions of Experiment B.

Figure 4 shows simulated values in yellow and experimental values in black. Despite their similarity, differences in temperature distribution can be seen between sensors 3 and 4.

Figure 5 shows the relationship between relative error and EAWHE length for Experiment B.

Based on simulation results, it can be inferred that the proposed model can generate air temperature values with an error of less than 6.20%. The relative error between the experimental and simulated temperatures is 3.14% for a length of 2.50 m, 4.33% for 3.40 m, 5.53% for 4.40 m, and 6.10% for 5.40 m. These results indicate a proportional increase in relative error with increasing EAWHE length. Table 3 summarizes the simulated values from experiments A and B. The table shows the temperature differences between the values, the relative errors generated in each simulation, and the seed values used to obtain the results.

TABLE III. SEEDS USED, TEMPERATURE DIFFERENCES, AND RELATIVE ERRORS OBTAINED IN EXPERIMENTS A AND B.

| Test setup | Seed | $\Delta T$ (°C) | Relative error (%) |
|---|---|---|---|
| Tin → S1 (2.50 m) → S4 | 5 | 0.36 | 1.25 |
| Tin → S1 (3.40 m) → S4 | 5 | 0.49 | 1.79 |
| Tin → S1 (4.40 m) → S4 | 5 | 0.64 | 2.40 |
| Tin → S1 (2.50 m) → S7 | 17 | 0.90 | 3.14 |
| Tin → S1 (3.40 m) → S7 | 5 | 1.19 | 4.33 |
| Tin → S1 (4.40 m) → S7 | 5 | 1.47 | 5.53 |
| Tin → S1 (5.40 m) → S7 | 5 | 1.57 | 6.10 |

The greatest errors occur when temperatures are simulated from Tin to the farthest sensor (S7). Conversely, the simulated temperature values between $T_{in}$ and S4 exhibit smaller relative errors and a smaller temperature difference ($\Delta T$). In most cases, using seed 5 yields lower temperature differences and relative errors between the simulated and measured temperature values. However, for the $T_{in}$-S1-S7 configuration, seed 17 yields the best results. The accuracy of the proposed model was evaluated by comparing the simulated temperature values with the temperature parameters recorded by the sensors. The RMSE value was lower (0.5096) when simulating conditions from $T_{in}$ to S4. However, for conditions from $T_{in}$ to S7, the RMSE value is 1.3088. Therefore, it can be assumed that the model provides more reliable predictions for conditions from Tin to S4.

## IV. Conclusions

The main contribution of this research lies in the development and rigorous validation of the DARL (Data of Air and Random Length) Model, establishing a new paradigm for predicting air temperature distribution in EAWHE. The model is positioned as a direct solution to the challenge posed by the high dependence on physical instrumentation (sensors) and the high computational costs associated with thermal characterization in field conditions.

The air temperature values obtained through simulation using the proposed model inside the EAWHE offer relative errors ranging from 1.25% to 6.10%. Given this range, it can be inferred that the model consistently produces results that align with the experimental data. However, the temperature values obtained through simulation at each tested position are not normally distributed. Therefore, position measures (quartile 3, quartile 2, quartile 1, and the interquartile range) should be used to characterize the magnitude. Future studies employing this method should use the Shapiro-Wilk test or another statistical test to determine the normality of the simulated parameters. Experiment A produces lower relative errors and a significantly lower root mean square error (RMSE) than Experiment B; therefore, researchers and geothermal development engineers are recommended to use Experiment A to minimize sensor usage and reduce the cost of acquiring these instruments. The validation demonstrates that model accuracy is inversely related to length, as the prediction error increases proportionally with the increase in the EAWHE extension. This limitation is evident in the comparison of the experimental configurations.

The methodological value and originality of the work focus on the implementation of an efficient technique. Unlike conventional models, which require complex geometric discretization of the pipe and soil to solve the heat transfer differential equations, the DARL model achieves its robustness through the integration of Fermat's prime numbers as seeds to generate PRN. This methodological choice creates a correlation criterion that, when interacting with experimental data, can efficiently replace detailed physical modeling. The other simulation techniques should be explored, and the quality of results obtained from different simulator seed values should be verified. Future results should be compared with those obtained in this research. The developed model should be implemented in geothermal heat exchangers under real operating conditions and in different geological and environmental settings to compare experimental and simulated results.